\documentclass[fleqn]{llncs}

\usepackage{amssymb}
\usepackage{graphicx}
\usepackage{asymptote}
\usepackage{amssymb}
\usepackage{float}
\usepackage{subfig}
\usepackage{amsmath}
\usepackage{pgf}
\usepackage{tabularx}
\usepackage{booktabs}

\usepackage{url}
\urldef{\mailsa}\path|ramamurt@iai.uni-bonn.de|
\urldef{\mailsb}\path|christian.bauckhage@iais.fraunhofer.de|
\urldef{\mailsc}\path|buza@iai.uni-bonn.de|

\graphicspath{{./images/}}
\begin{document}

\mainmatter  

\title{Using Echo State Networks for Cryptography}

\titlerunning{Using Echo State Networks for Cryptography}

%
%
\author{R. Ramamurthy \and C. Bauckhage \and K. Buza \and S. Wrobel}
\authorrunning{R. Ramamurthy et al.}

\institute{Department of Computer Science, University of Bonn, Bonn, Germany}

%
%

\toctitle{Echo State Networks for Data Encryption}
\tocauthor{}
\maketitle

\begin{abstract}
Echo state networks are simple recurrent neural networks that are easy to implement and train. Despite their simplicity, they show a form of memory and can predict or regenerate sequences of data. We make use of this property to realize a novel neural cryptography scheme. The key idea is to assume that Alice and Bob share a copy of an echo state network. If Alice trains her copy to memorize a message, she can communicate the trained part of the network to Bob who plugs it into his copy to regenerate the message. Considering a byte-level representation of in- and output, the technique applies to arbitrary types of data (texts, images, audio files, etc.) and practical experiments reveal it to satisfy the fundamental cryptographic properties of diffusion and confusion.
\end{abstract}

\section{Introduction}
\label{sec:intro}

The emerging field of \textit{neural cryptography} is a sub-field of cryptography that deals with artificial neural networks for encryption and cryptanalysis. 

Early contributions in this area considered cryptographic systems based on recursive auto encoders and showed that feed-forward networks trained via back-propagation can encrypt plain-text messages in the activation patterns of hidden layer neurons \cite{clark1998neural}. Later work introduced key-exchange systems where coupled neural networks synchronize to establish common secret keys \cite{kanter2002secure}; while the original approach was not completely secure \cite{Klimov2002}, more recent work showed that modern convolutional interacting neural networks can indeed learn to protect their communication against adversary eaves-droppers \cite{abadi2016learning}. Another popular idea is to combine chaotic dynamics and neural networks \cite{li2005chosen,Lian20091296,Wang2010,Yu2006333,zhou2004novel}. For example, chaotic neural networks were used for image encryption and experimentally verified to be secure and chaotic Hopfield networks were found to be able to generate random binary sequences for text encryption. 

Given this short survey, the novel idea for neural cryptography proposed in this paper can be seen as a hybrid approach that harnesses chaotic dynamics and the deterministic outcome of a training procedure. Namely, we propose to use echo state networks \cite{jaeger2001echo} both for encryption and decryption. 

Considering the classic  scenario where Alice and Bob exchange messages and want to protect their communication against Eve's eavesdropping, we assume that both share an identical copy of an echo state network whose internal states evolve according to a non-linear dynamical system. To encrypt a message (a text, an image, etc.), Alice feeds it into her copy of the network and trains the output weights such that the network reproduces the input. She then sends these output weights to Bob who uses them to run his copy of the network which will regenerate the message. Eve, on the other hand, may retrieve the communicated output weights, but without the corresponding echo state network (its structure, input weights, and internal weights), she will not be able to decipher the message. Our experiments with this kind of private-key or symmetric cryptography system reveal the approach to be easy to use, efficient, scalable, and secure. 

Next, we briefly summarize the basic theory behind echo state networks and how to use them as auto encoders that memorize their input. We then discuss how to harness them for cryptography and present experiments which underline that our approach satisfies the fundamental cryptographic properties of diffusion and confusion.

\section{Echo State Networks as Memories}
\label{sec:esn}

Echo state networks (ESNs) follow the paradigm of reservoir computing where a large reservoir of recurrently interconnected neurons processes sequential data. The central idea is to randomly generate weights $\boldsymbol{W}^{i} \in \mathbb{R}^{n_{r} \times n_{i}}$ between input and reservoir neurons  as well as weights $\boldsymbol{W}^{r} \in \mathbb{R}^{n_{r} \times n_{r}}$ between reservoir neurons. Only  the weights $\boldsymbol{W}^{o} \in \mathbb{R}^{n_{o} \times n_{r}}$  between reservoir and output neurons are trained in order to adapt the network to a particular task.

At time $t$, the states of the input, output, and reservoir neurons are collected in $\boldsymbol{x}_t 	\in \mathbb{R}^{n_{i}}$ , $\boldsymbol{y}_t \in \mathbb{R}^{n_{o}}$, and $\boldsymbol{r}_t \in \mathbb{R}^{n_{r}}$, respectively, and their evolution over time is governed by the following non-linear dynamical system
\begin{align}
\boldsymbol{r}_t & = (1-\alpha)  \boldsymbol{r}_{t-1} + \alpha \, f_r \bigl( \boldsymbol{W}^{r} \boldsymbol{r}_{t-1} + \boldsymbol{W}^{i} \boldsymbol{x}_t \bigr) \label{eq:update} \\
\boldsymbol{y}_t & =  f_o \bigl( \boldsymbol{W}^{o} \boldsymbol{r}_t \bigr) \label{eq:predict}
\end{align}
where  $\alpha \in [0, 1]$ is called the leaking rate. The function $f_r(\cdot)$ is understood to act component-wise on its argument and is typically a sigmoidal activation function. For the output layer, however, $f_o(\cdot)$ is usually just a linear or softmax function depending on the application context. 

To train an echo state network, one provides a training sequence of input data $\boldsymbol{x}_1 , \boldsymbol{x}_2 , \ldots, \boldsymbol{x}_T$ gathered in a matrix $\boldsymbol{X} \in \mathbb{R}^{n_{i} \times T}$ together with a sequence of desired outputs $\boldsymbol{y}_1, \boldsymbol{y}_2, \ldots, \boldsymbol{y}_T$ gathered in $\boldsymbol{Y} \in \mathbb{R}^{n_{o} \times T}$.  The training sequence is fed into the network and the internal activations that result from iterating equation \eqref{eq:update} are recorded in a matrix $\boldsymbol{R}=[\boldsymbol{r}_1,\boldsymbol{r}_2, \ldots, \boldsymbol{r}_T] \in \mathbb{R}^{n_{r} \times T}$. Appropriate output weights $\boldsymbol{W}^{o}$ can then be determined using least squares 
\begin{equation}\label{eq:train}
\boldsymbol{W}^{o} = \boldsymbol{Y} \boldsymbol{R}^T (\boldsymbol{R}\boldsymbol{R}^T + \beta \boldsymbol{I})^{-1}
\end{equation}
where $\beta$ is a regularization constant. However, for a good practical performance, the scale $a$ of $\boldsymbol{W}^{i}$ and the spectral radius $\rho$ of $\boldsymbol{W}^{r}$ have to be chosen carefully. Together with the leaking rate $\alpha$, these parameters are rather task specific, yet, useful, commonly adhered to general guidelines are given in \cite{lukovsevivcius2012practical}.

Because of its recurrent connections, the reservoir of an echo state network can be understood as a non-linear high-dimensional expansion of the input data that has a memory of the past. The temporal reach of this memory is called ``memory capacity'' and bounded by the number of reservoir neurons \cite{jaeger2002short}. An entire input sequence (e.g.~a text file) can therefore be stored in- and retrieved from the reservoir provided the reservoir is large enough. Hence, our idea in this paper is to produce an echo state network with a large reservoir and to train it to memorize an input sequence. Once the training is complete, we let the network run freely to (re)generate the memorized sequence.

\section{ESN-Based Encryption and Decryption}
\label{sec:scheme}

We consider the classic cryptographic scenario where Alice and Bob want to secure their communication against Eve's eavesdropping.
Using a \textit{secret key}, Alice converts her messages known as \textit{plaintexts} into encrypted messages known as \textit{ciphertexts}. She then sends the cyphertexts to Bob who uses the same key to convert them back into plaintexts. 

Given this setup, our idea is to ``memorize'' a given message using an echo state network at one end of a communication channel and to ``recall'' it at the other end using the same network. If Alice and Bob share an identical copy of the network, Alice can train it to memorize the data and transmits only the resulting weights $\boldsymbol{W}^{o}$ over the insecure channel. Bob then plugs these weights into his copy of the network and runs it to reconstruct Alice's message. In other words, the weight matrices $\boldsymbol{W}^{i}$ and $\boldsymbol{W}^{r}$ and leaking rate $\alpha$ of the echo state network constitute the secret key of our cryptographic system.  Without it Eve can not decipher the transmitted cyphertext $\boldsymbol{W}^{o}$.

\subsection{Representing Data}

In our practical implementations of the above scheme, we consider byte-level representations of messages. This allows for flexibility and wide applicability because, in the memory of a computer, texts or images are represented as a byte-stream after all. To further increase flexibility, we consider a ``one hot'' encoding of individual bytes where each of the 256 possible values is represented as a 256-dimensional binary vector.

\subsection{Memorizing Data}\label{memorize}

Given any byte sequence $\boldsymbol{B}=[b_1, b_2, \ldots, b_N]$ of input data, we train and apply an echo state network as follows: First, we append a dummy byte $b_0$ at the beginning of the original sequence $\boldsymbol{B}$ so as to make the later recall process independent of the value of the original first byte in the sequence. Second, we encode the resulting sequence to obtain $\boldsymbol{H} = [\boldsymbol{h}_0, \boldsymbol{h}_1, \ldots, \boldsymbol{h}_N]$ where each $\boldsymbol{h}_i$ is a binary vector of length 256. Given $\boldsymbol{H}$, we then set the in- and output sequence for an echo state network to 
\begin{align}
\boldsymbol{X} & = [\boldsymbol{h}_0, \boldsymbol{h}_1, \ldots, \boldsymbol{h}_{N-1}] \\
\boldsymbol{Y} & = [\boldsymbol{h}_1, \boldsymbol{h}_2, \ldots, \boldsymbol{h}_{N}]
\end{align}
where the indices of the vectors in sequences $\boldsymbol{X}$ and $\boldsymbol{Y}$ differ by one time step. Given an echo state network with input weights $\boldsymbol{W}^{i}$ and reservoir weights $\boldsymbol{W}^{r}$, we then iterate the system in \eqref{eq:update} and \eqref{eq:predict} and learn appropriate output weights $\boldsymbol{W}^{o}$ according to \eqref{eq:train}.

\subsection{Recalling Data}\label{recall}

Once $\boldsymbol{W}_{o}$ has been determined, it can be plugged into an identical copy of the echo state network at the other end of a communication channel. This network can then regenerate the encoded message one element at a time. To this end, we consider the dummy byte $b_0$ and ``one hot'' encode it to obtain $\boldsymbol{x}_0 = \boldsymbol{h}_0$. Using this as the first input to the network, we run the system in \eqref{eq:update} and \eqref{eq:predict} to obtain $\boldsymbol{y_t}$ from $\boldsymbol{x}_t$. At each time step, we consider the network output $\boldsymbol{y_t}$, which is not necessarily a binary vector, as a vector of probabilities for different bytes. We thus subject it to the softmax function which returns a 1 for the most likely entry and 0s for all others. The resulting binary vector is then used as the input $\boldsymbol{x_{t+1}}$ for the next iteration of the network. Moreover, we decode the binary vectors obtained in each iteration into bytes $b_t$ and collect them in a matrix $\boldsymbol{S}$, which is exactly the original sequence $\boldsymbol{B}$ memorized by the echo state network.

\subsection{Working with ``Data Chunks''}\label{subsec: chunks}

As the size $N$ of data sequence increases, the size $n_{r} \in \mathcal{O}(N)$ of a reservoir that can memorize it increases, too. This makes the matrix multiplications $\boldsymbol{W}^{r} \boldsymbol{r}_t$ required for the network's state updates expensive. In fact, the total cost for $N$ internal updates will be of order $\mathcal{O}(N^3)$ and, to reduce this cost, we adopt a ``divide-and-conquer'' strategy where we split the data into chunks of size $m$ and employ a small reservoir to memorize each chunk at an effort of $\mathcal{O}(m^3)$. Hence, for an entire sequence, i.e.~for $\tfrac{N}{m}$ chunks, efforts reduce to $\mathcal{O}(\tfrac{N}{m} \times m^3) = \mathcal{O}(Nm^2)$.


\section{Experiments and Results}\label{sec:experiments}

In our practical experiments, we found that echo state networks used as described above can indeed memorize and perfectly recall different types of data such as texts, images, audio files, videos, archives, etc. In this section we report results obtained from different kinds of security analysis of our cryptographic scheme. The parametrization of the echo state networks considered in these experiments is summarized in Tab.~\ref{experiments}.

\begin{table}[t!]
	\centering
	\caption{Echo state network configuration}
	\begin{tabularx}{1.0\textwidth}{l l}
		\toprule
		parameter & value \\
		\midrule
		chunk size $m$    & $200$ (reservoir size $n_{r}$ chosen as $0.95 \times m$)\\
		leaking rate $\alpha$    & $0.07$ \\
		spectral radius $\rho$ of $W^{r}$       & $1.0$\\
		input scaling $a$ of $W^{i}$       & $0.5$\\
		random seed & randomly chosen \\
		input connectivity & input neurons are connected to 30 \% of reservoir neurons \\
		reservoir connectivity & reservoir neurons are connected to 30\% of reservoir neurons\\
		activation function $f$ & logistic for the reservoir and softmax for the output\\
		\bottomrule
	\end{tabularx}
	\label{experiments}
\end{table}

\subsection{Security analysis}

Any cryptography system should be robust against common types of attacks such as brute force attacks, known-plaintext attacks, and ciphertext-only attacks. A brute force attack is an attack in which an attacker attempts to find the keys of the system through trial and error. It is evident from the Tab.~\ref{experiments} that the key space of our proposed system is very large and most of the parameters are unbounded. This renders brute force attacks extremely time consuming and practically infeasible.

\begin{figure}[t!]
	\centering
	\subfloat[]{\includegraphics[width=0.24\textwidth]{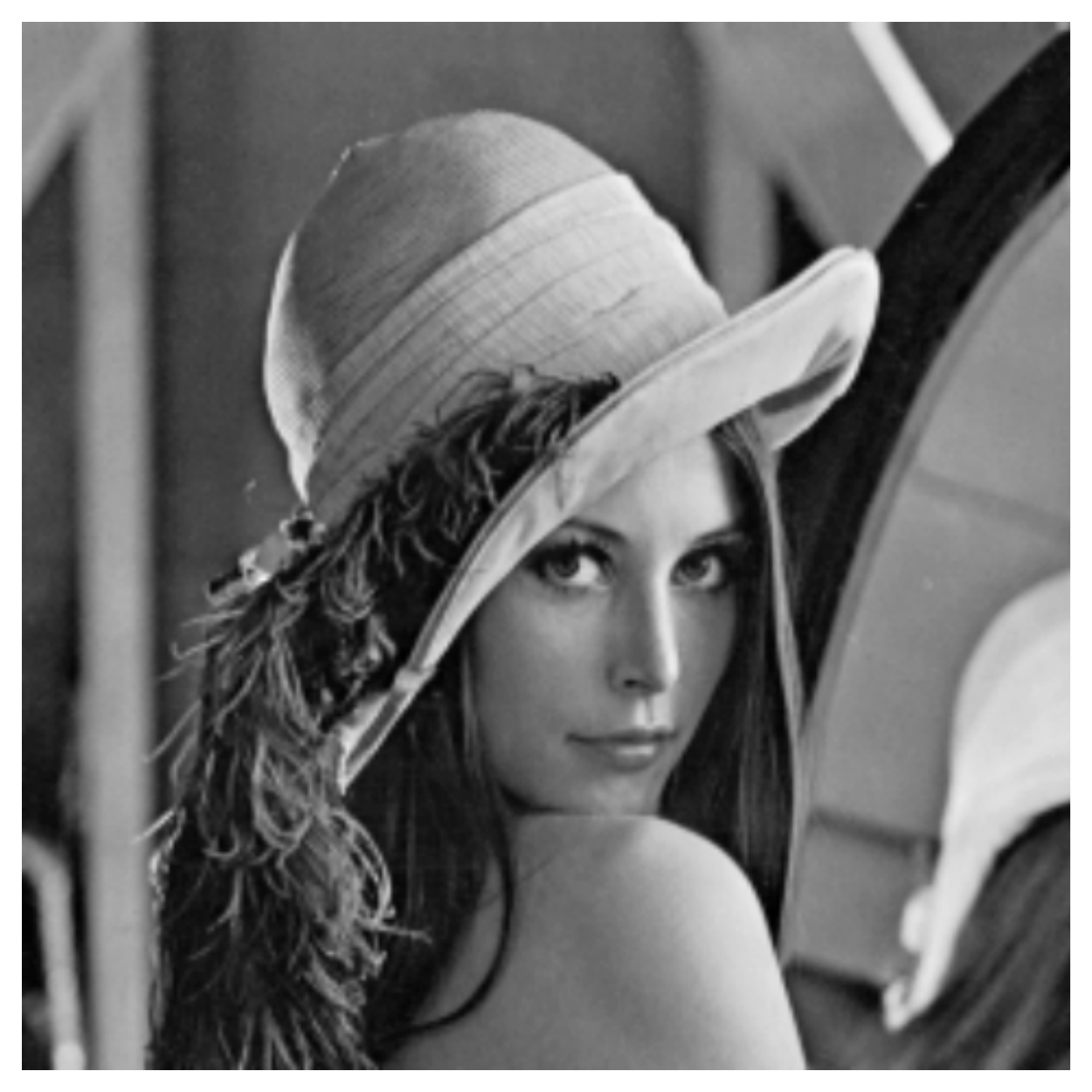}} \hfill
	\subfloat[]{\includegraphics[width=0.24\textwidth]{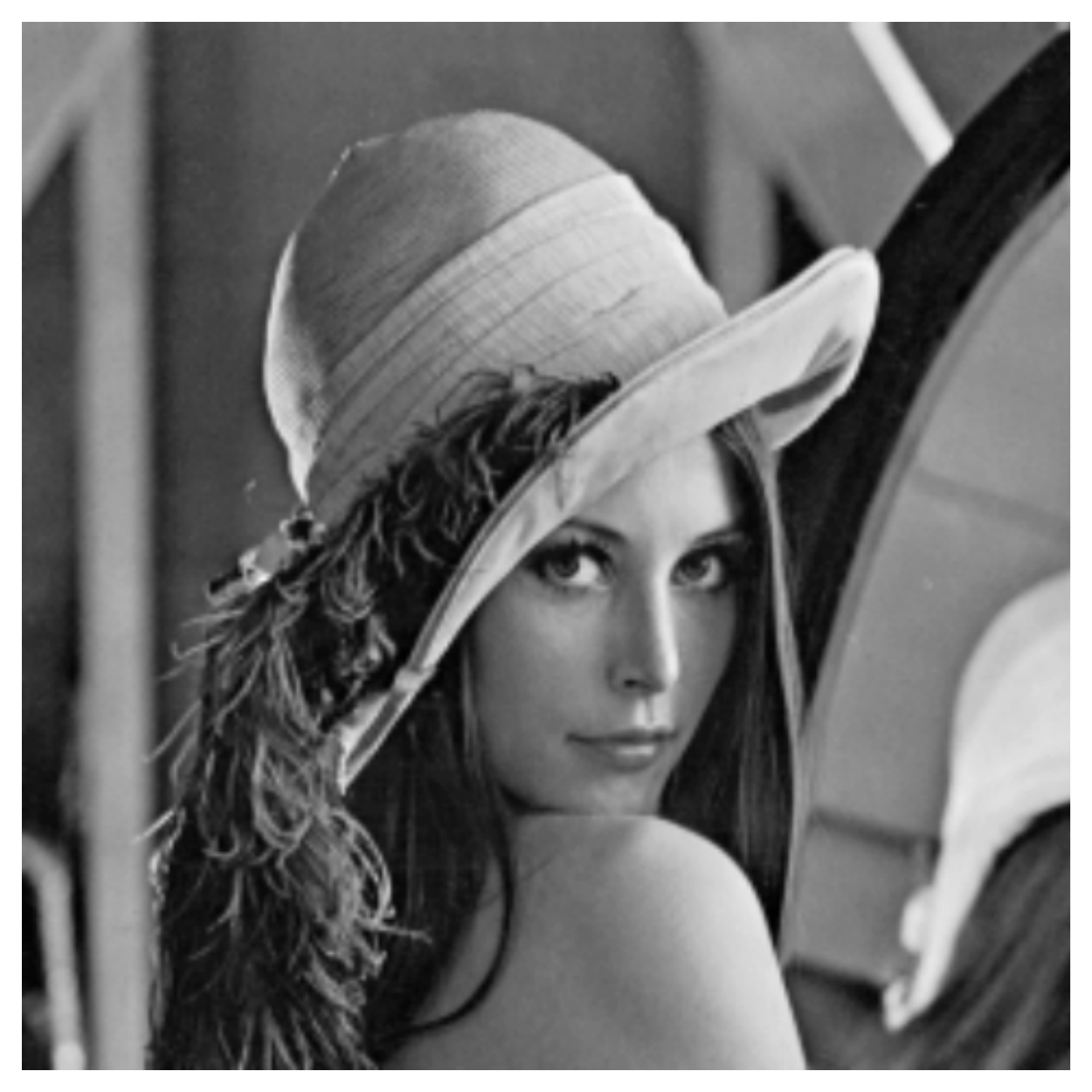}} \hfill 
	\subfloat[]{\includegraphics[width=0.24\textwidth]{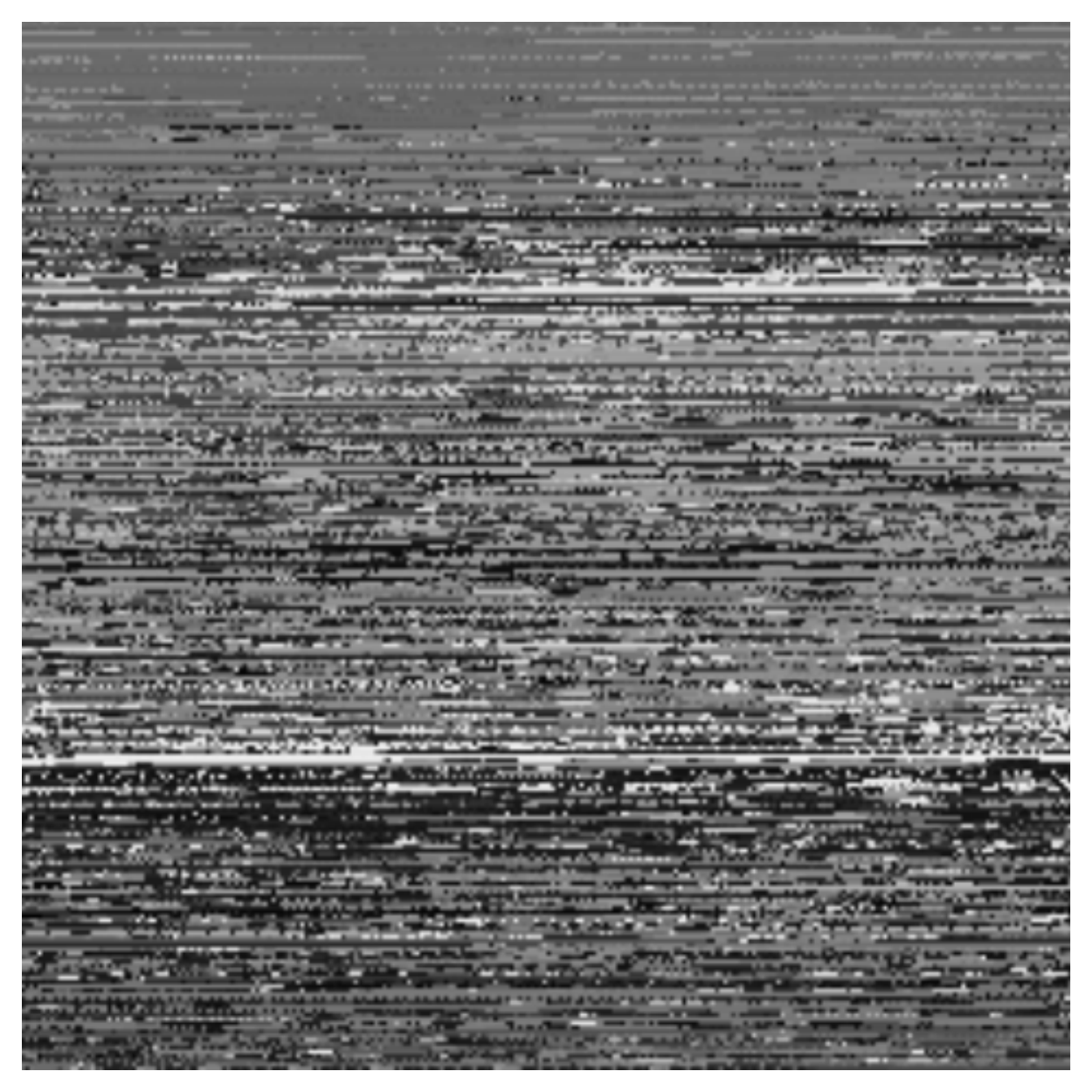}} \hfill
	\subfloat[]{\includegraphics[width=0.24\textwidth]{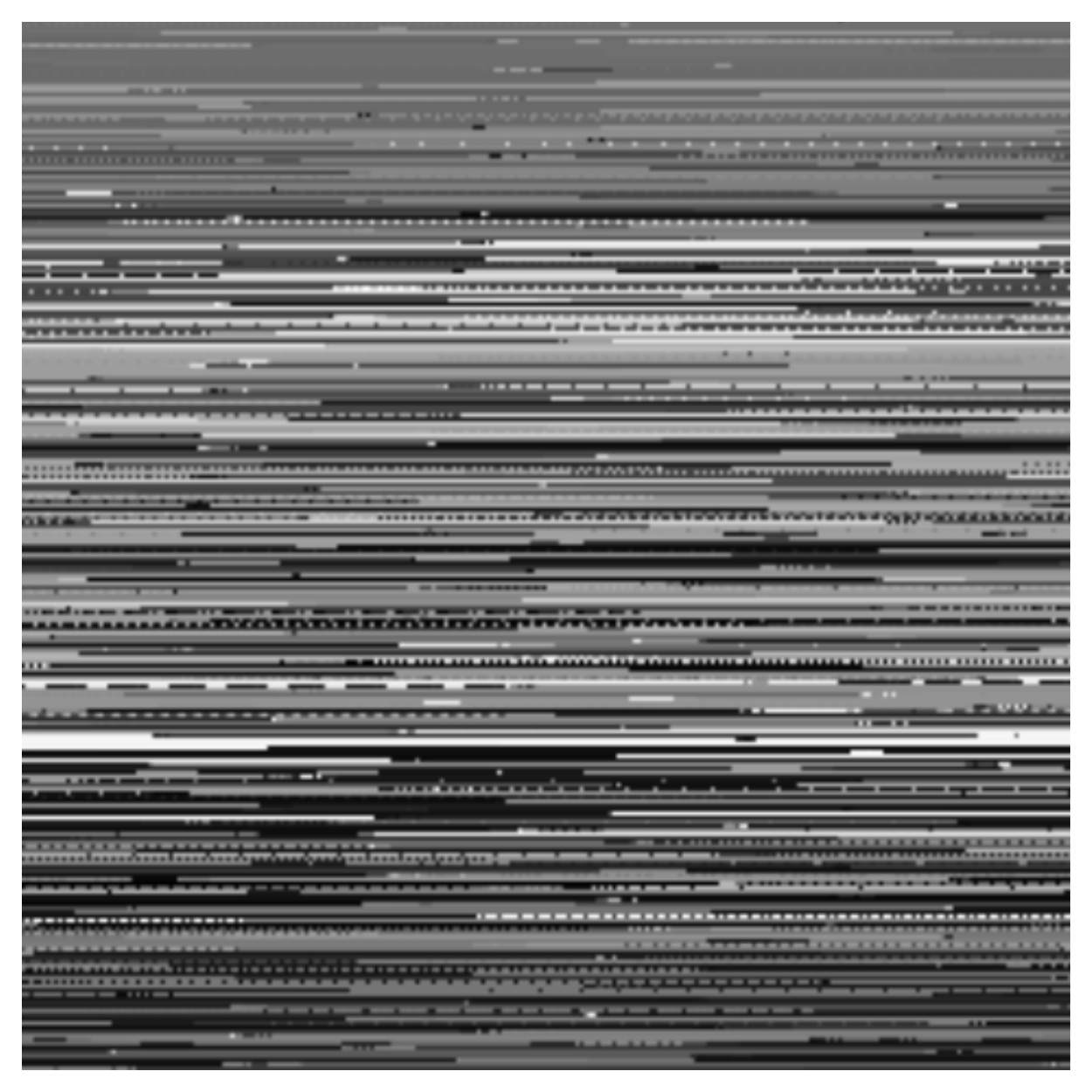}}
	
    \subfloat[]{\includegraphics[width=0.24\textwidth]{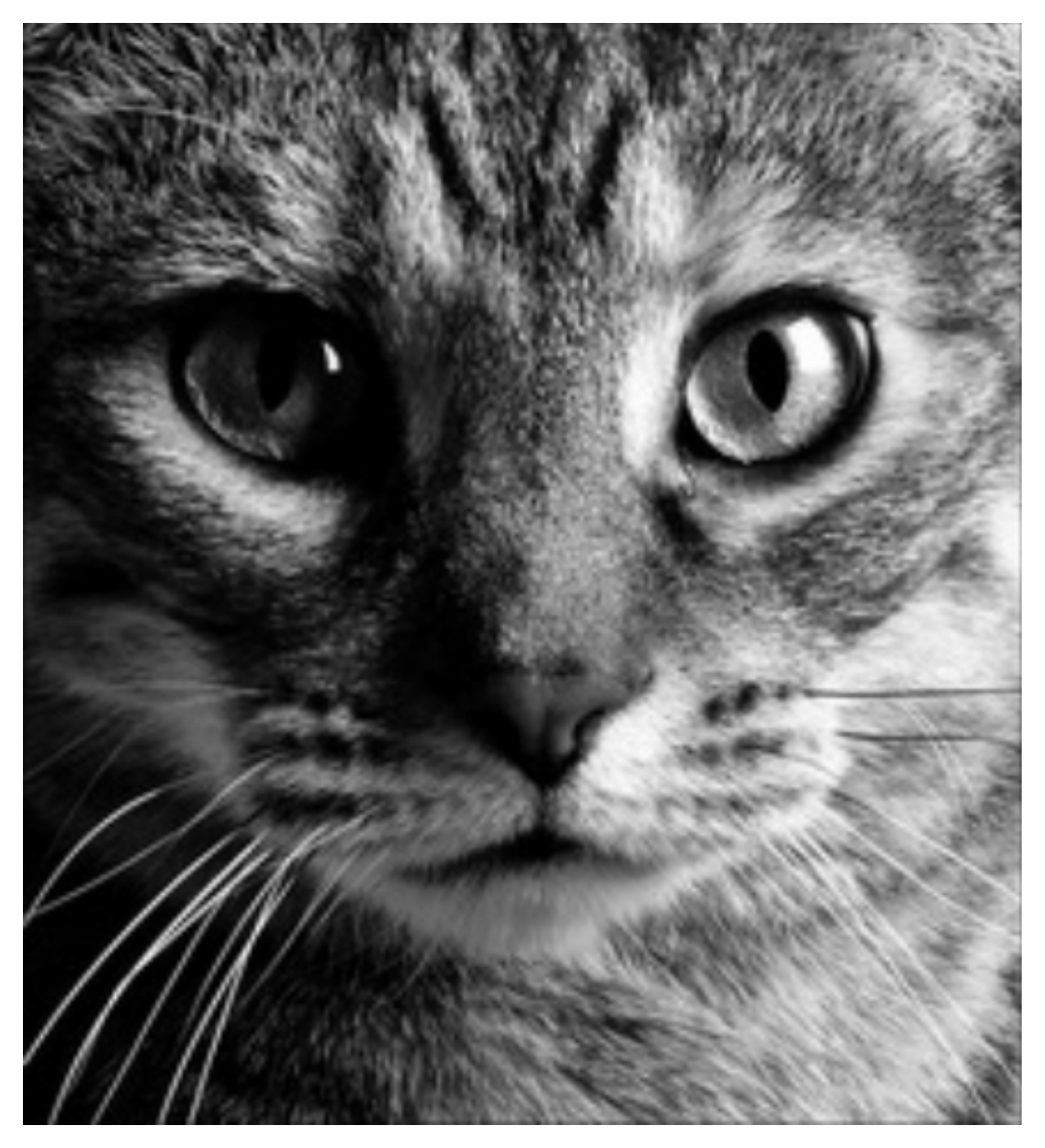}} \hfill
	\subfloat[]{\includegraphics[width=0.24\textwidth]{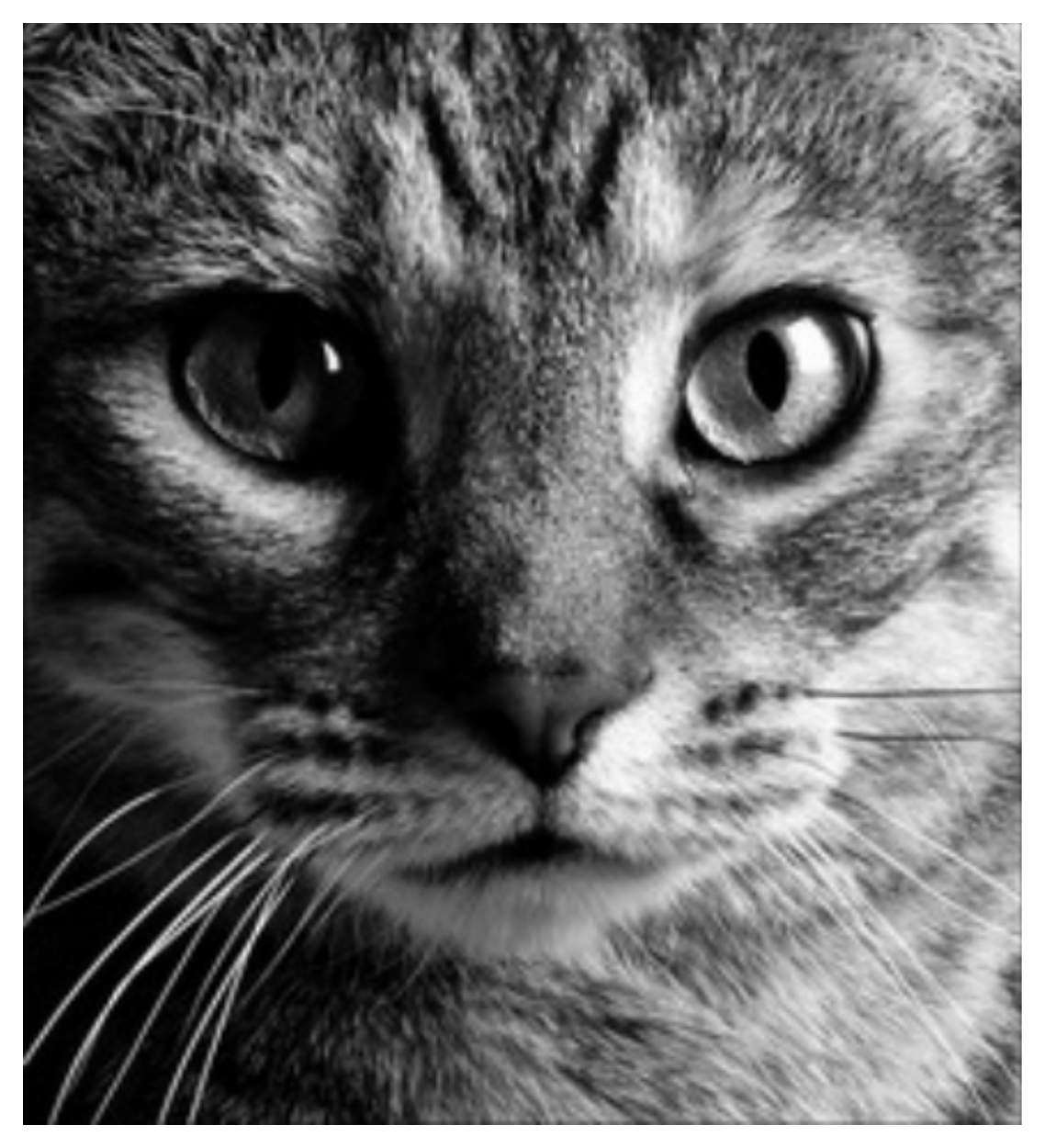}} \hfill 
	\subfloat[]{\includegraphics[width=0.24\textwidth]{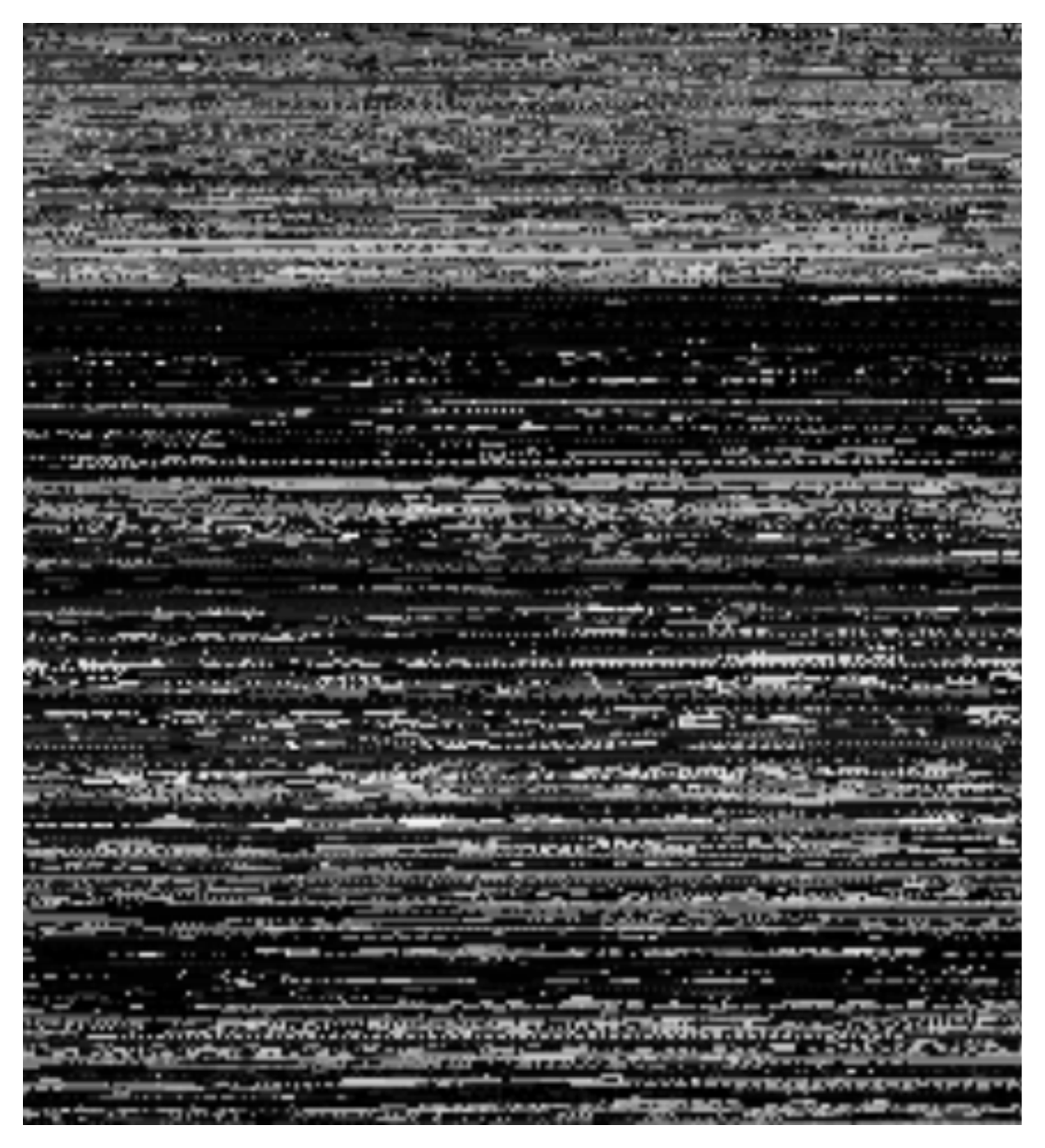}} \hfill
	\subfloat[]{\includegraphics[width=0.24\textwidth]{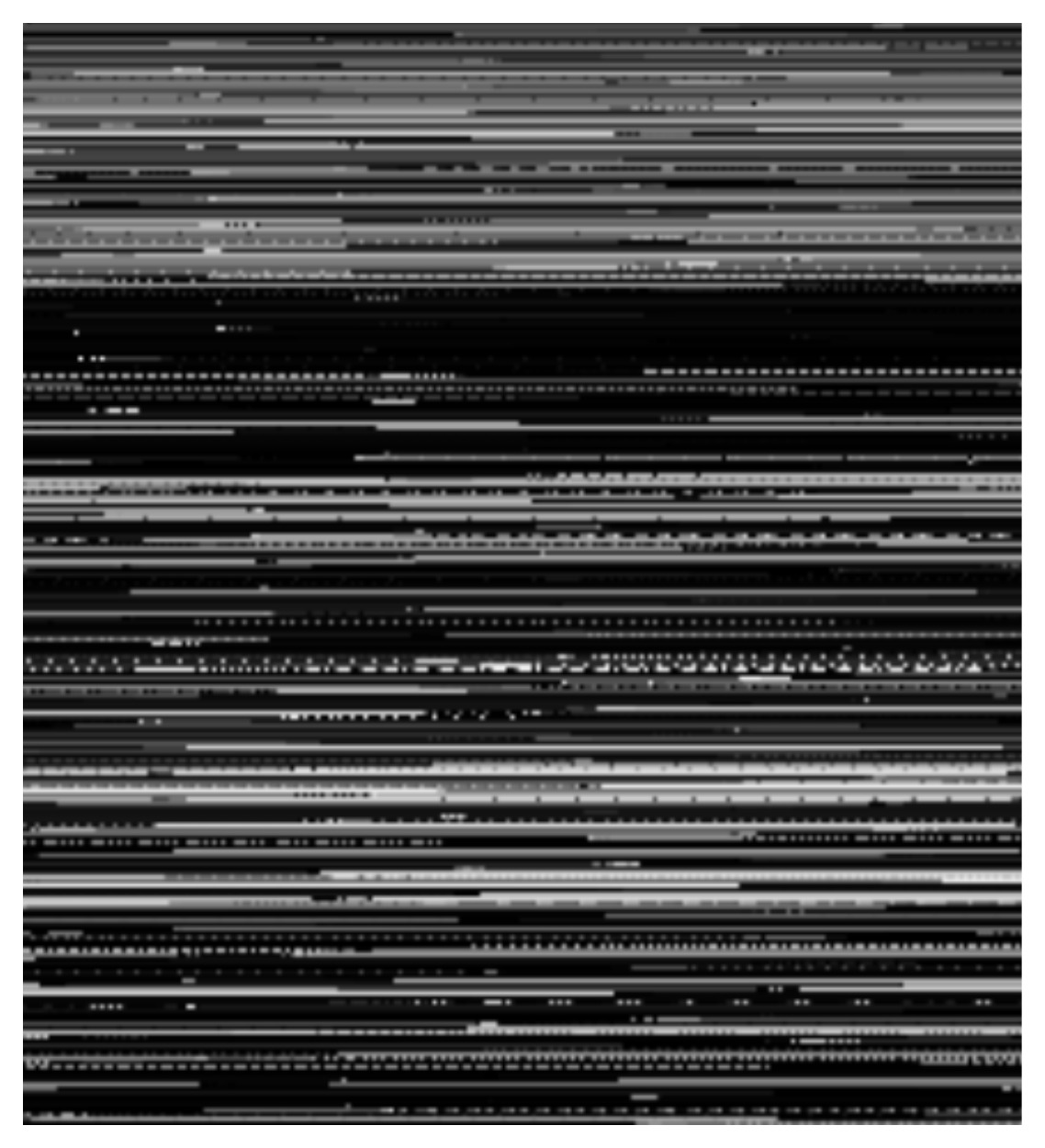}}	
	\caption{\textbf{Key sensitivity}: (a),(e) original images;  (b),(f) decrypted images using the same key (echo state network) as used for encryption; both decrypted images are identical to the originals;  (c),(d),(g),(h) decrypted images using slightly modified keys, i.e.~slightly modified echo state networks; here, all decrypted images differ considerably from the original images.}
	\label{key_sensitivity}
\end{figure}

Figures~\ref{key_sensitivity}(a) and (e) show two original images one of which (Lena) was given as a tiff file, the other (cat) as a png file. Both were encrypted and decrypted using the same echo state network. Decryption produced the images in Fig.~\ref{key_sensitivity}(b) and (f) which are identical to the original ones. However, when decrypting with networks with slightly modified parameters, i.e.~when using slight variations of the secret key, we obtained useless images as shown in Fig.~\ref{key_sensitivity}(c), (d), (g), and (h).
These results are prototypical and show that the system is highly sensitive to the secret key. This makes it robust against brute force attacks because decryption is only possible if all the parameters of the secret key are set precisely. 

\begin{figure}[t!]
	\centering
	\subfloat[]{\includegraphics[width=0.19\textwidth]{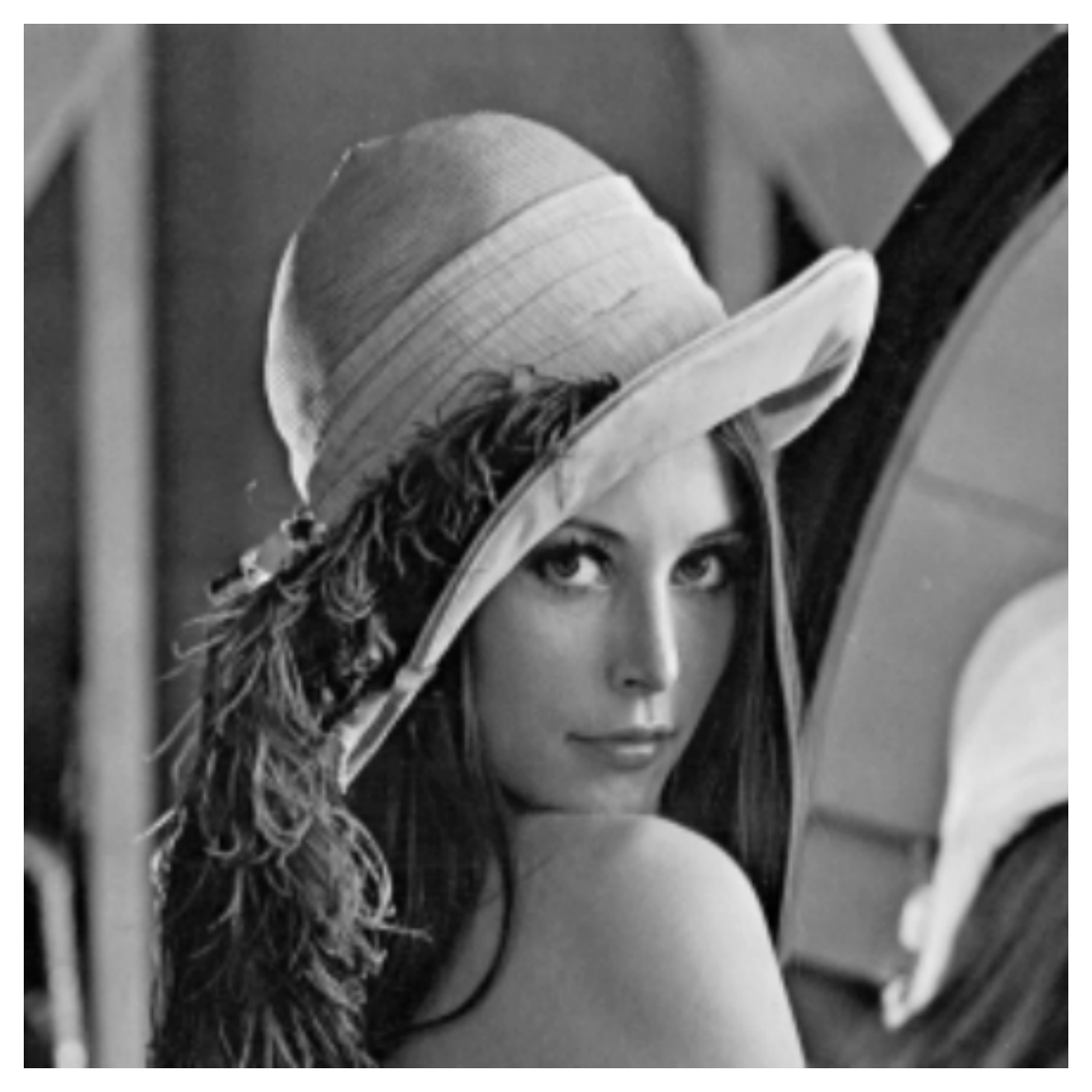}} \hfill
	\subfloat[]{\includegraphics[width=0.19\textwidth]{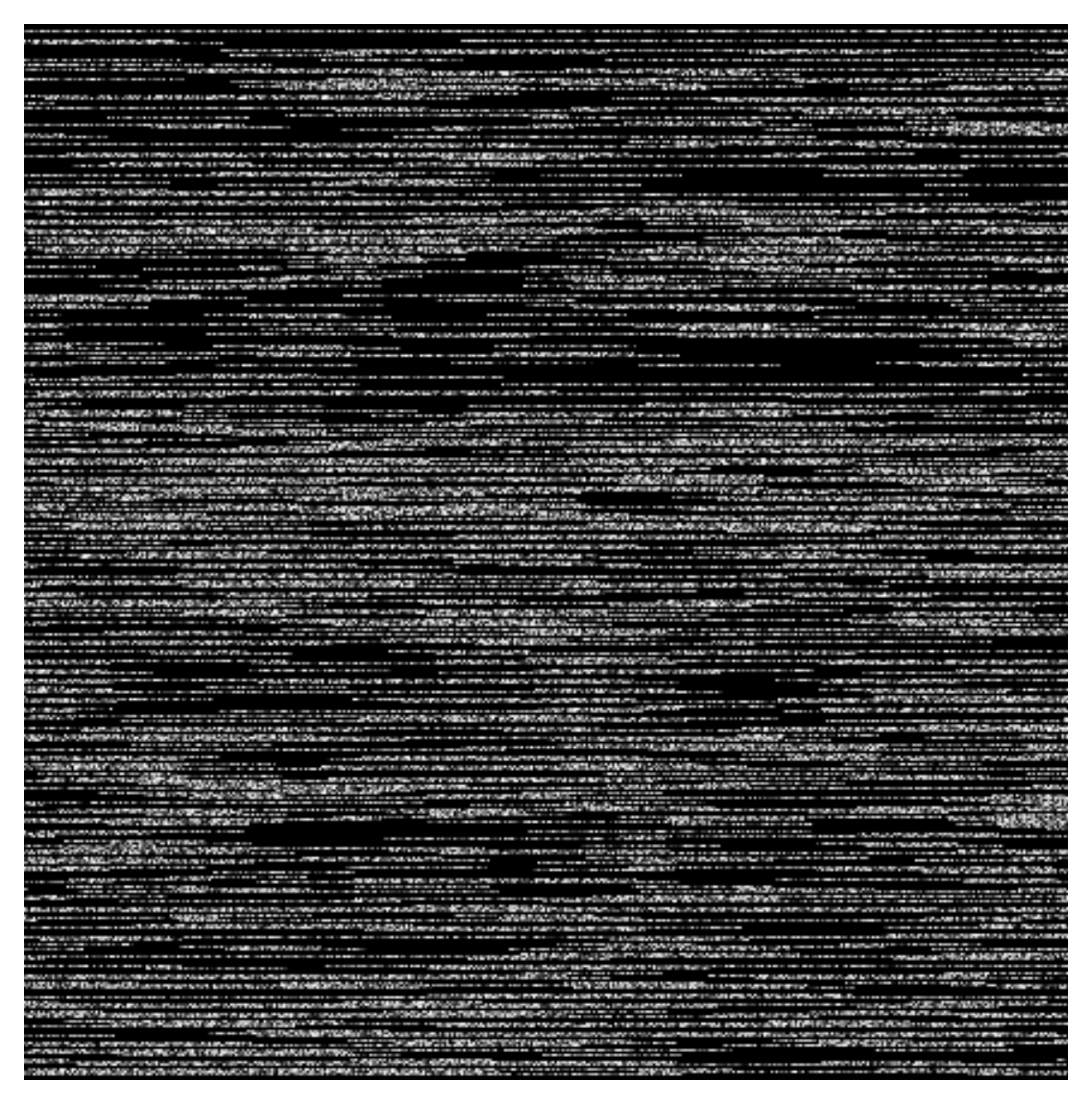}} \hfill
	\subfloat[]{\includegraphics[width=0.19\textwidth]{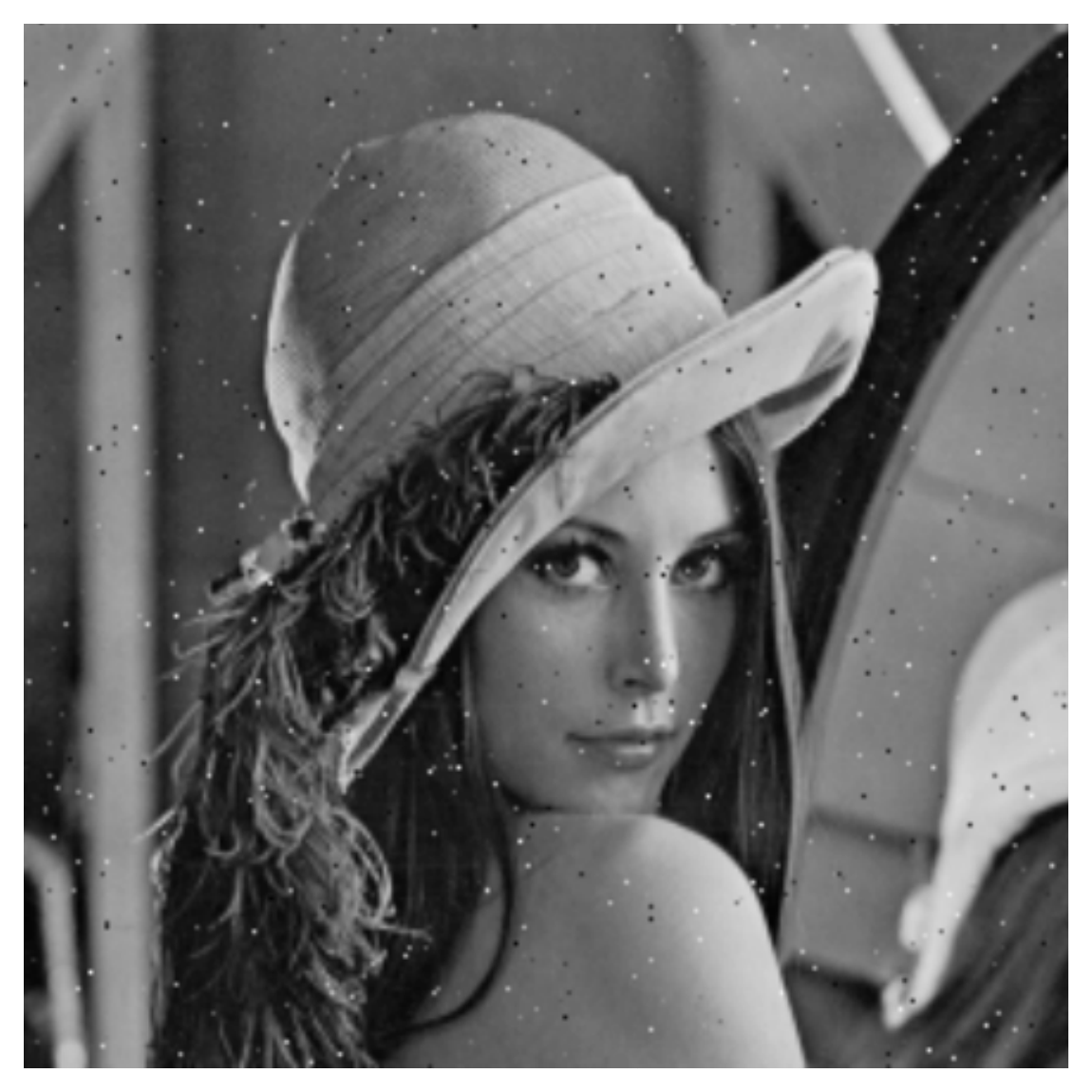}} \hfill
	\subfloat[]{\includegraphics[width=0.19\textwidth]{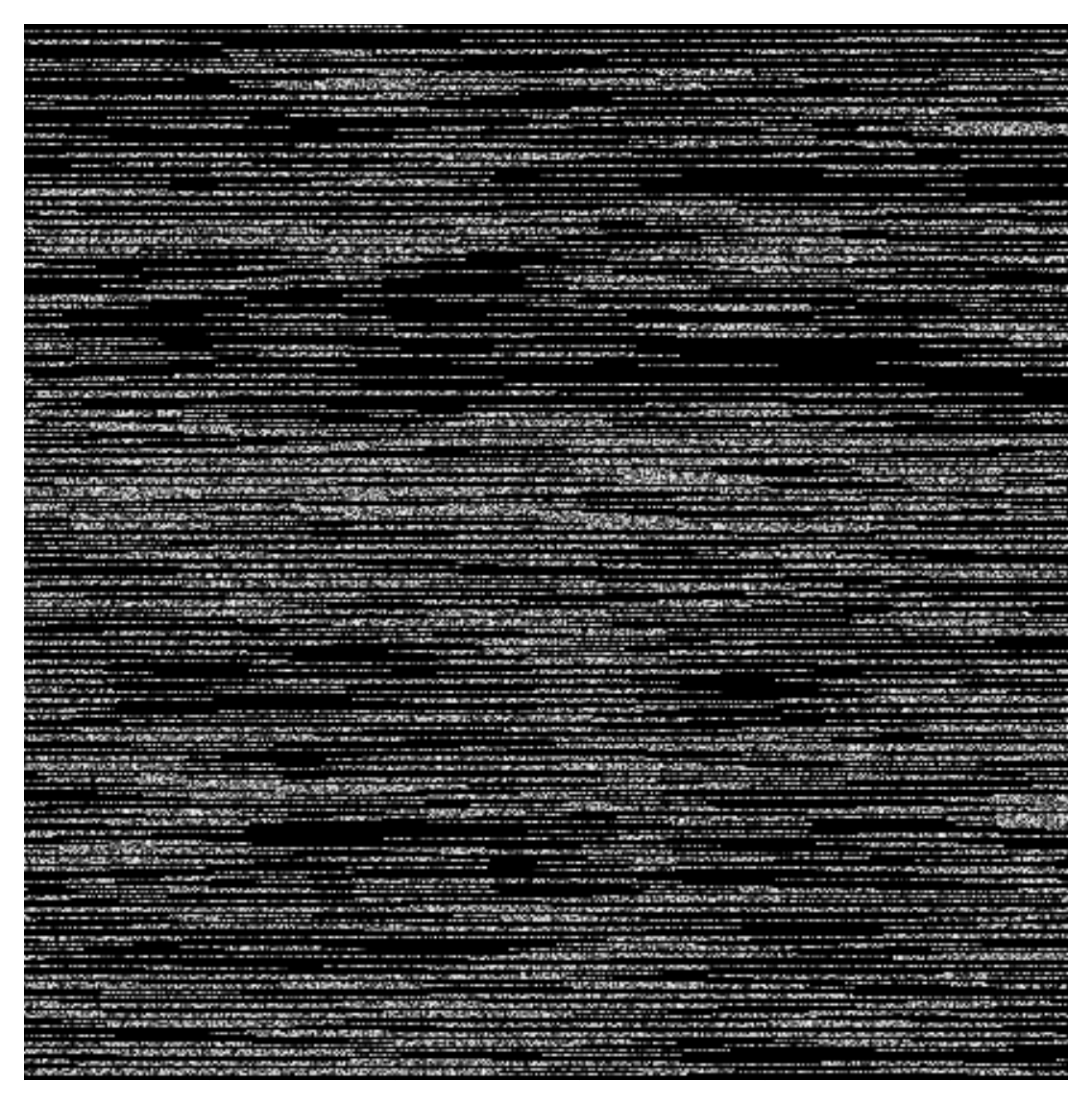}} \hfill
	\subfloat[]{\includegraphics[width=0.19\textwidth]{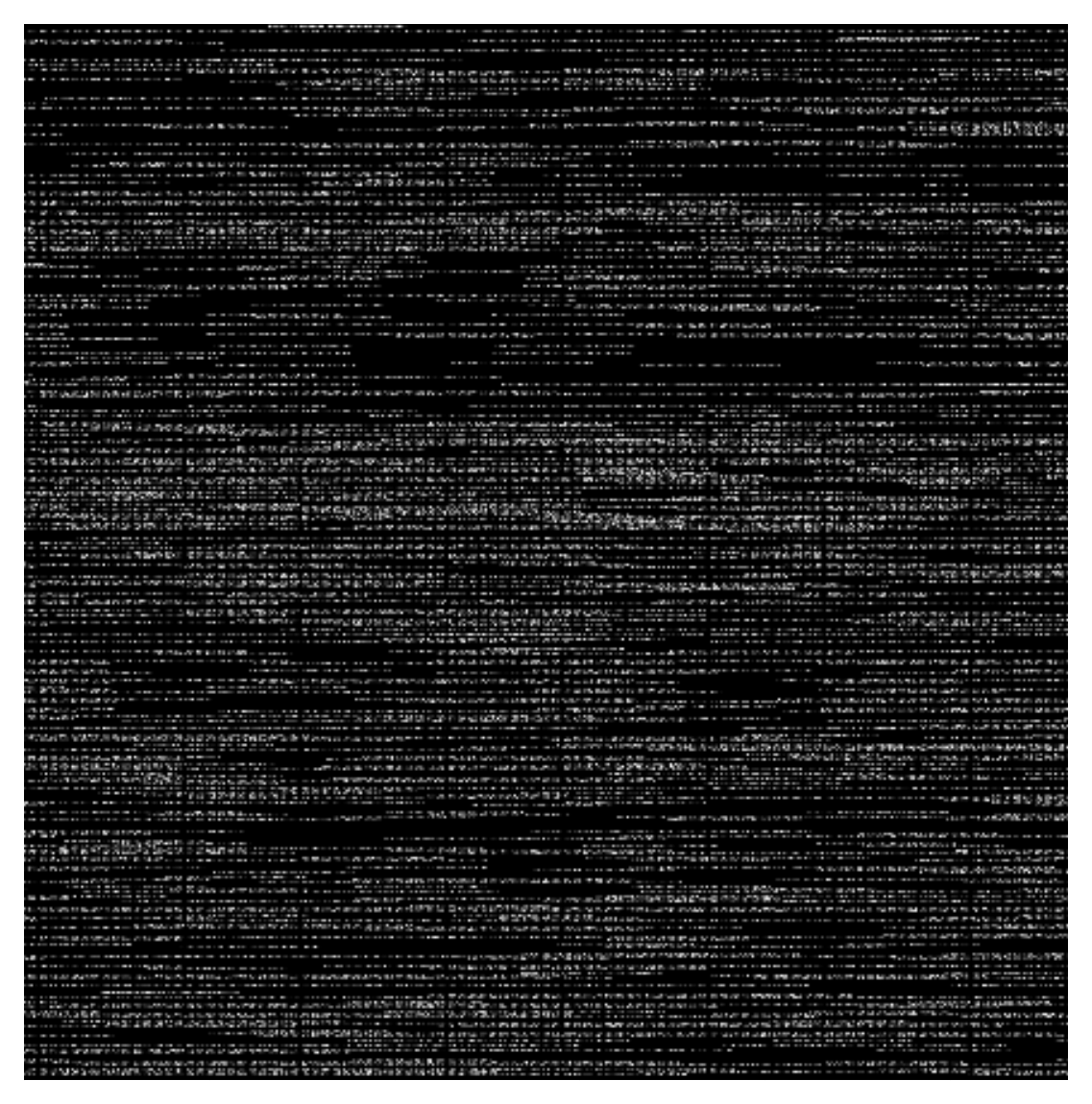}}
	
	\subfloat[]{\includegraphics[width=0.19\textwidth]{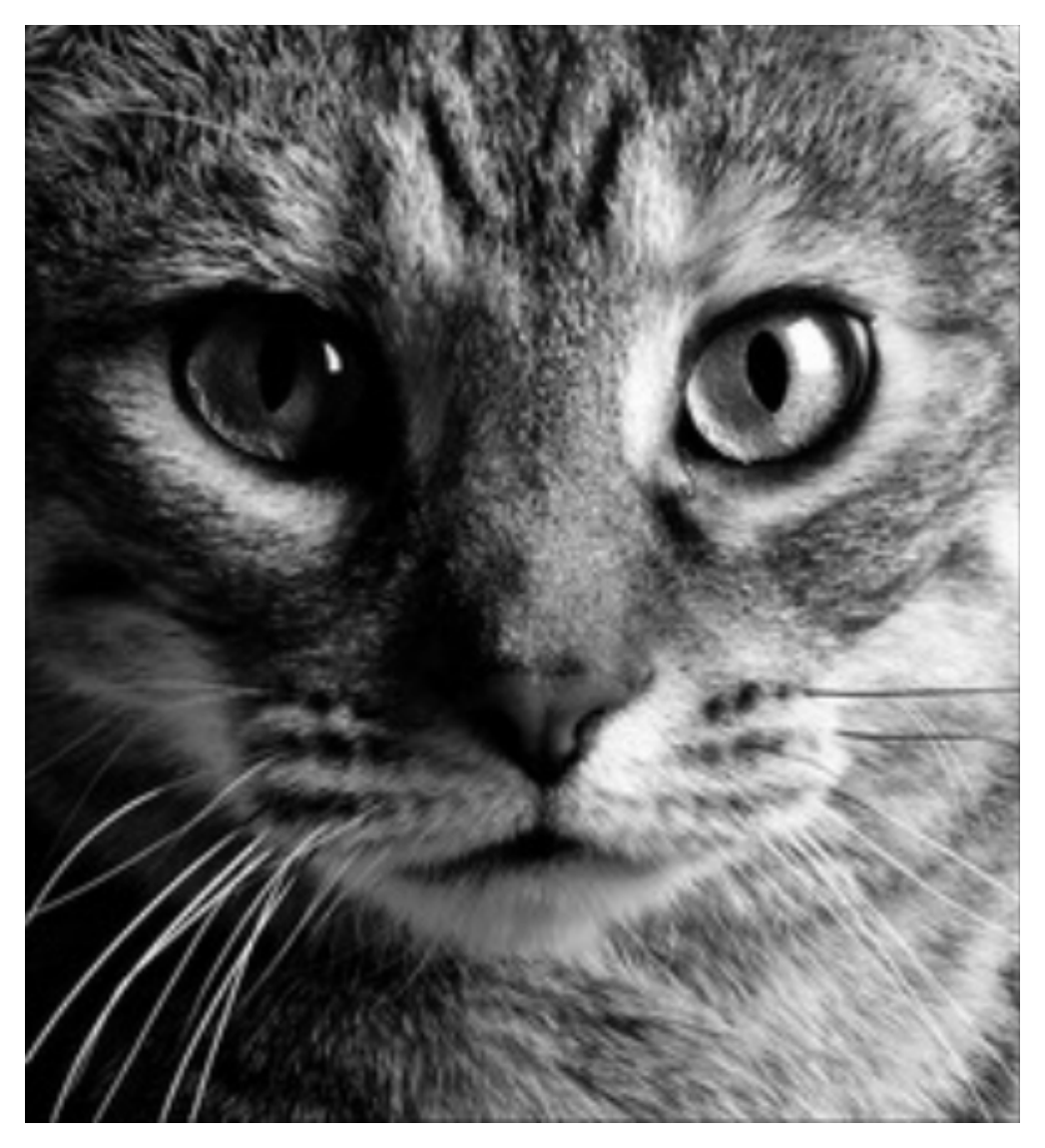}} \hfill
	\subfloat[]{\includegraphics[width=0.19\textwidth]{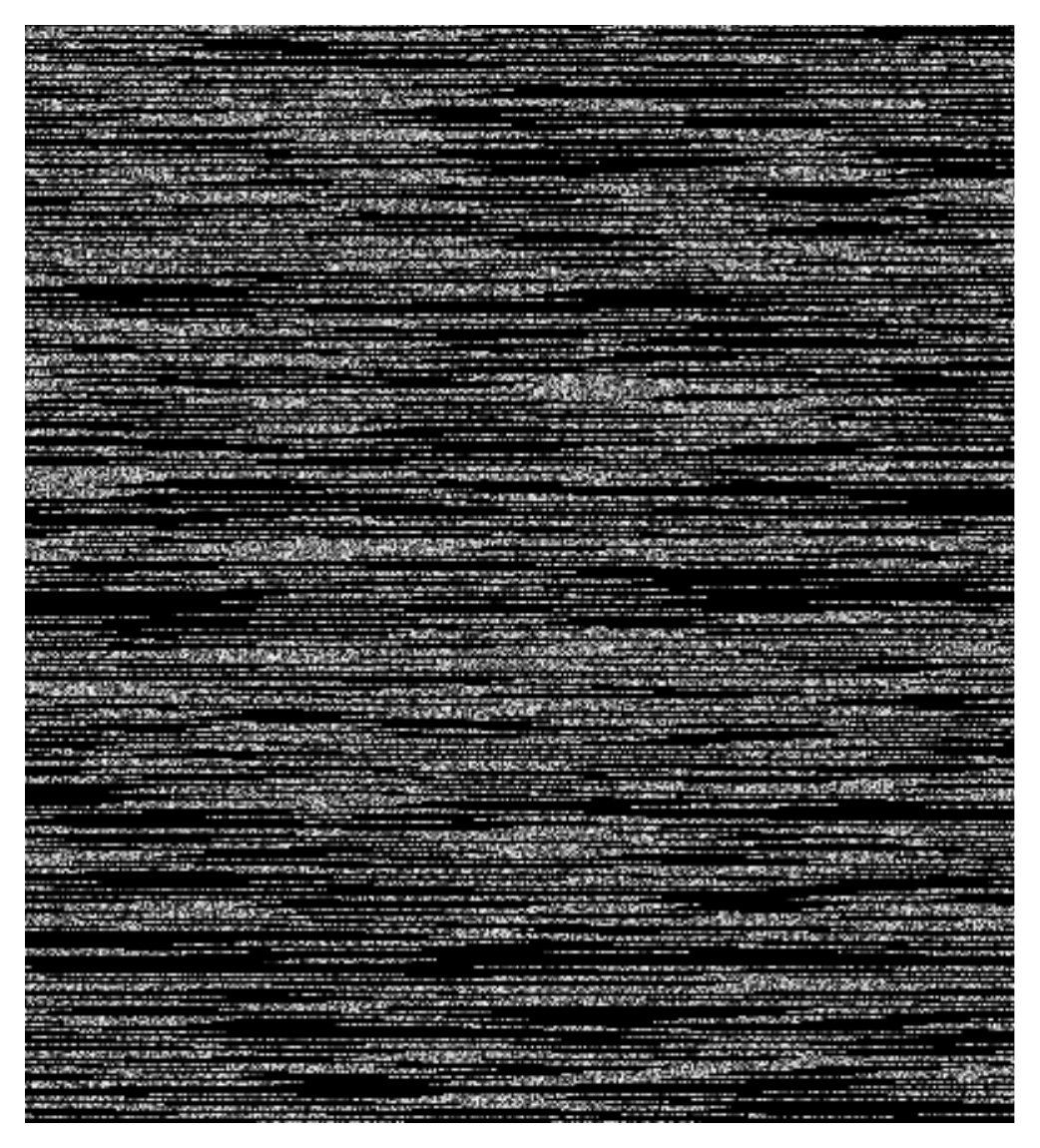}} \hfill
	\subfloat[]{\includegraphics[width=0.19\textwidth]{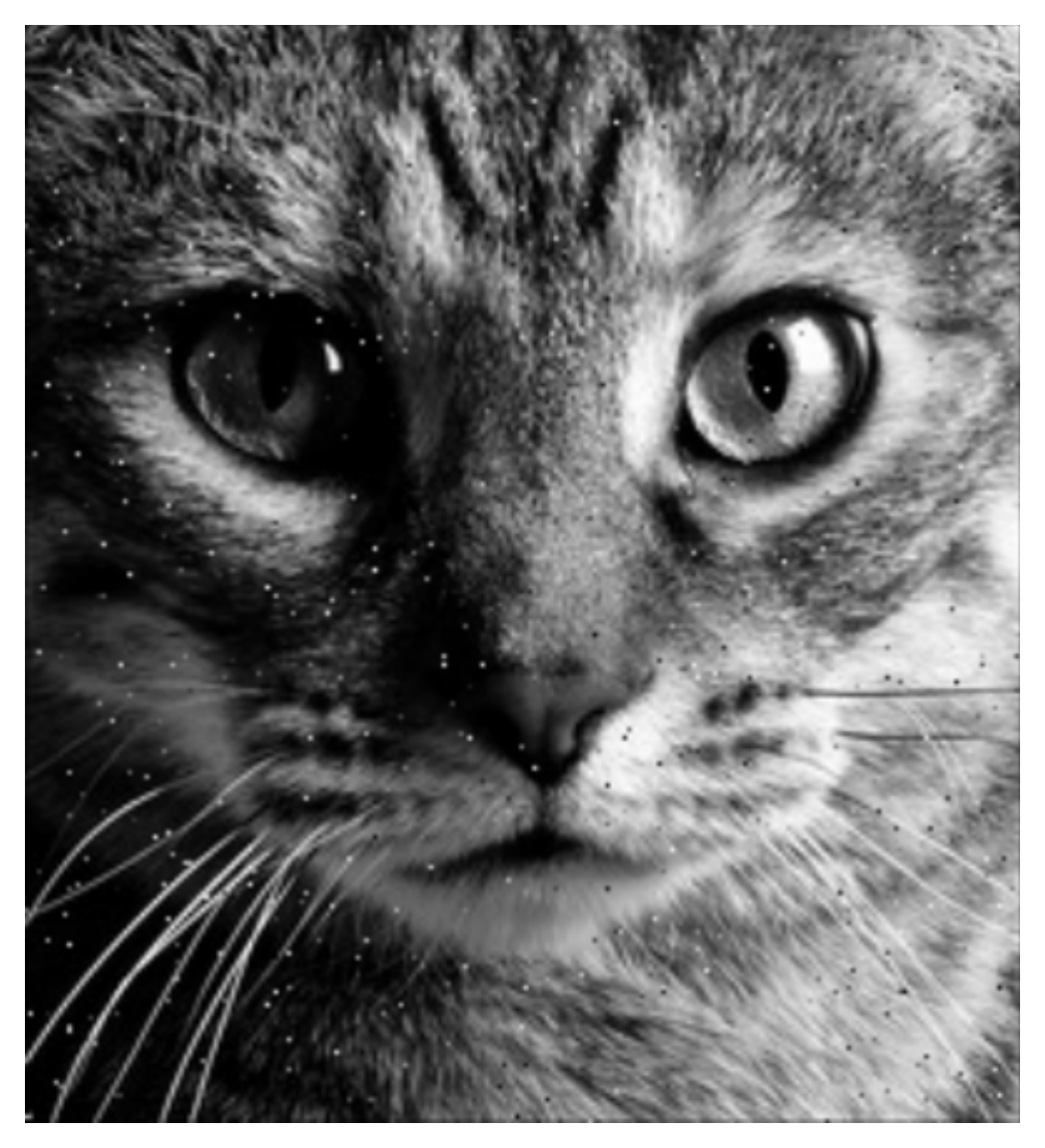}} \hfill
	\subfloat[]{\includegraphics[width=0.19\textwidth]{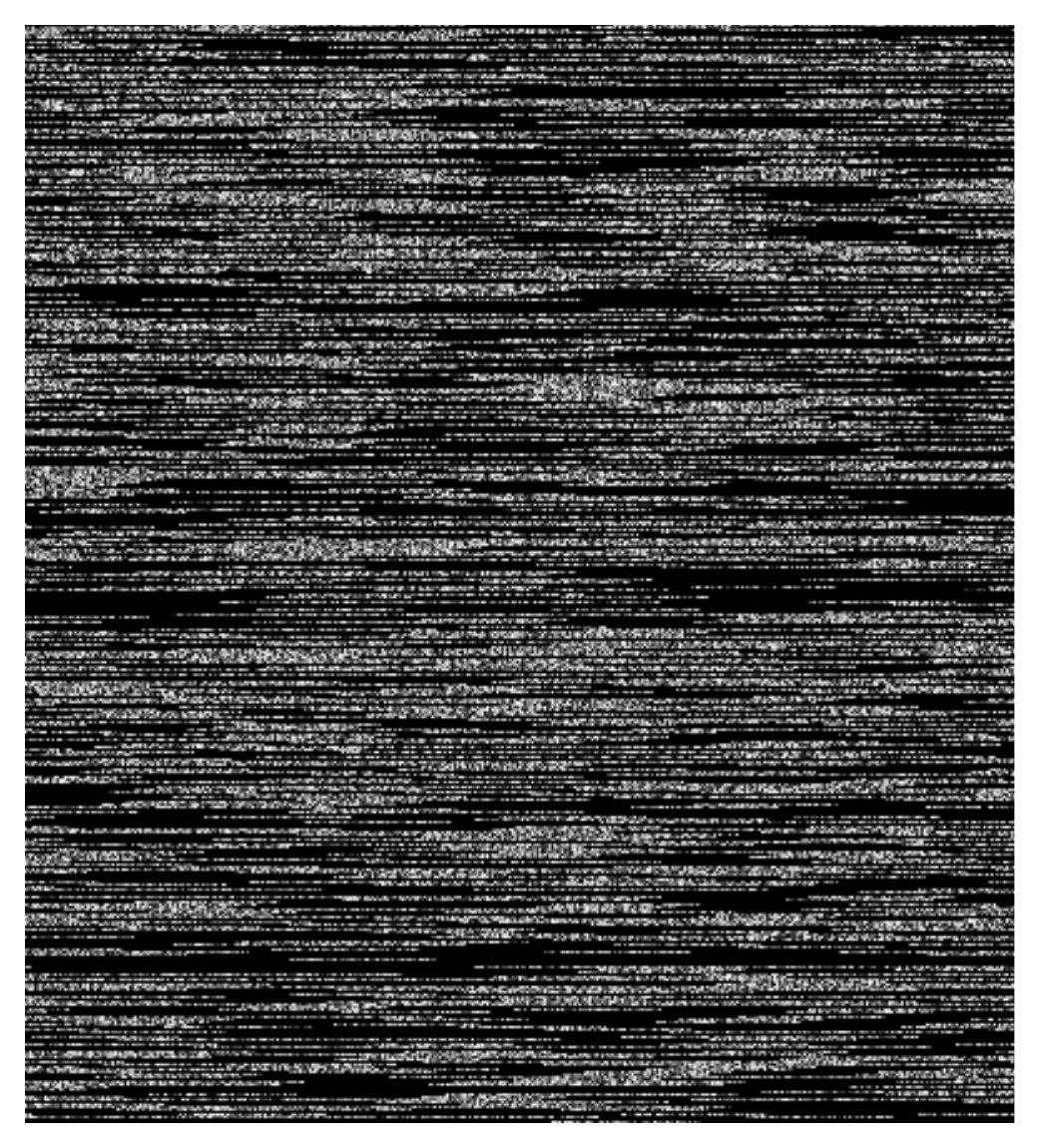}} \hfill
	\subfloat[]{\includegraphics[width=0.19\textwidth]{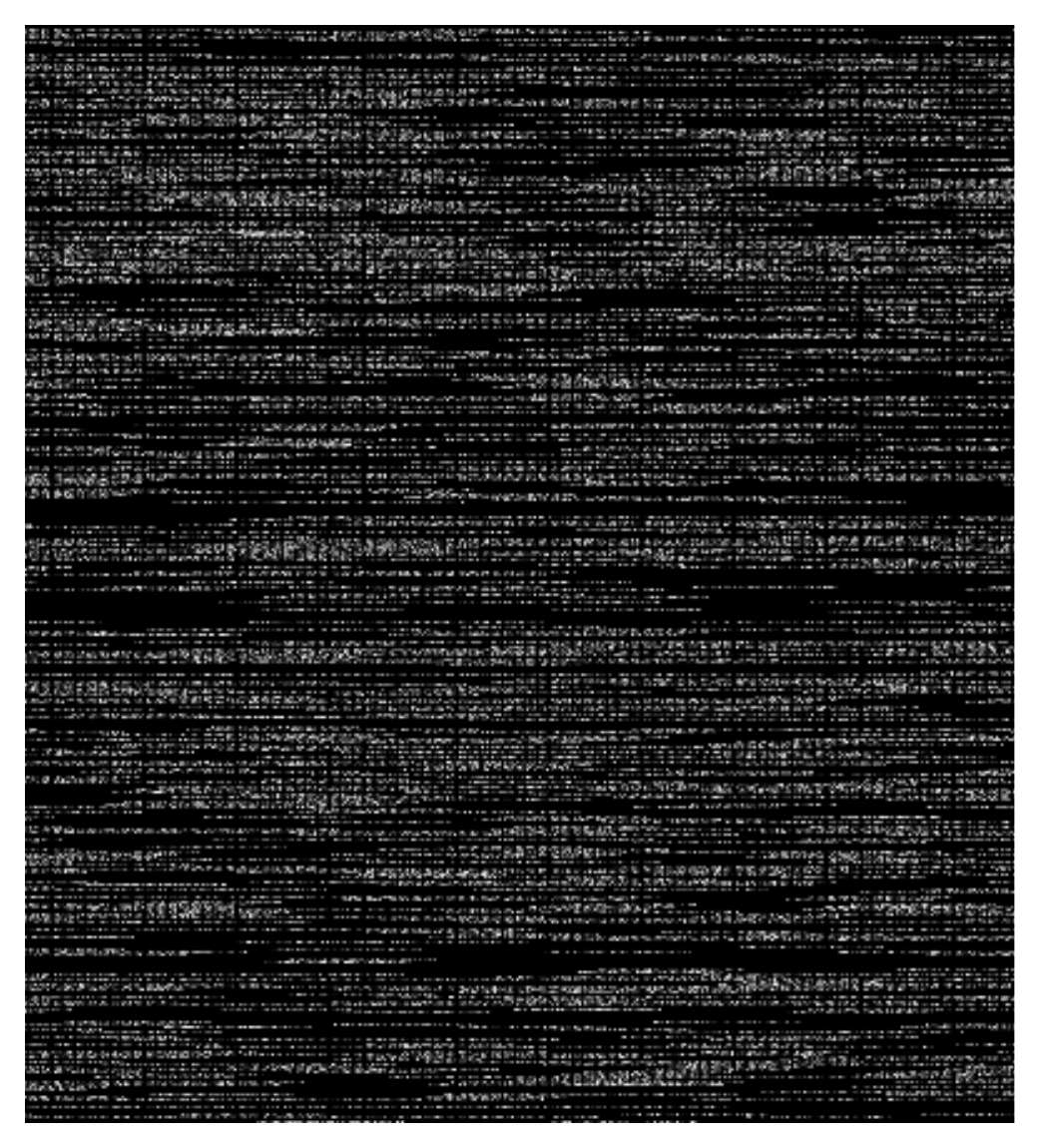}}
	\caption{\textbf{Plaintext sensitivity}: (a),(f) original images; (b),(g) encrypted images; (c),(h) original images with $1\%$ of their pixels randomly distorted; (d),(i) encryptions of the modified images; (e),(j) difference between encrypted original and encrypted modified images ($33.22\%$ and $37.78\%$, respectively).}
	\label{plaintext_sensitivity}
\end{figure}

\begin{figure}[t!]
	\centering
	\subfloat[]{\includegraphics[width=0.22\textwidth]{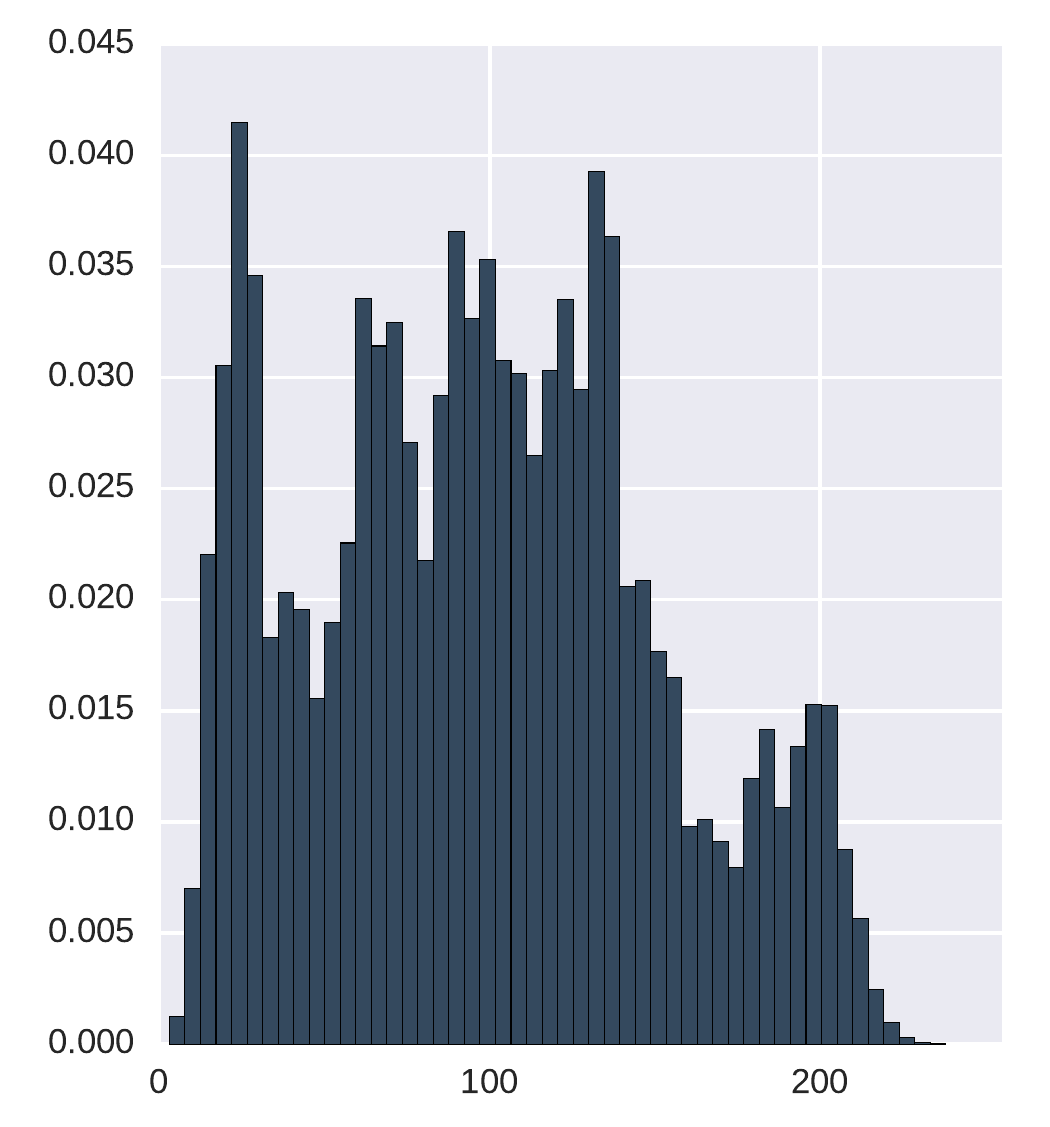}} \hfill
	\subfloat[]{\includegraphics[width=0.22\textwidth]{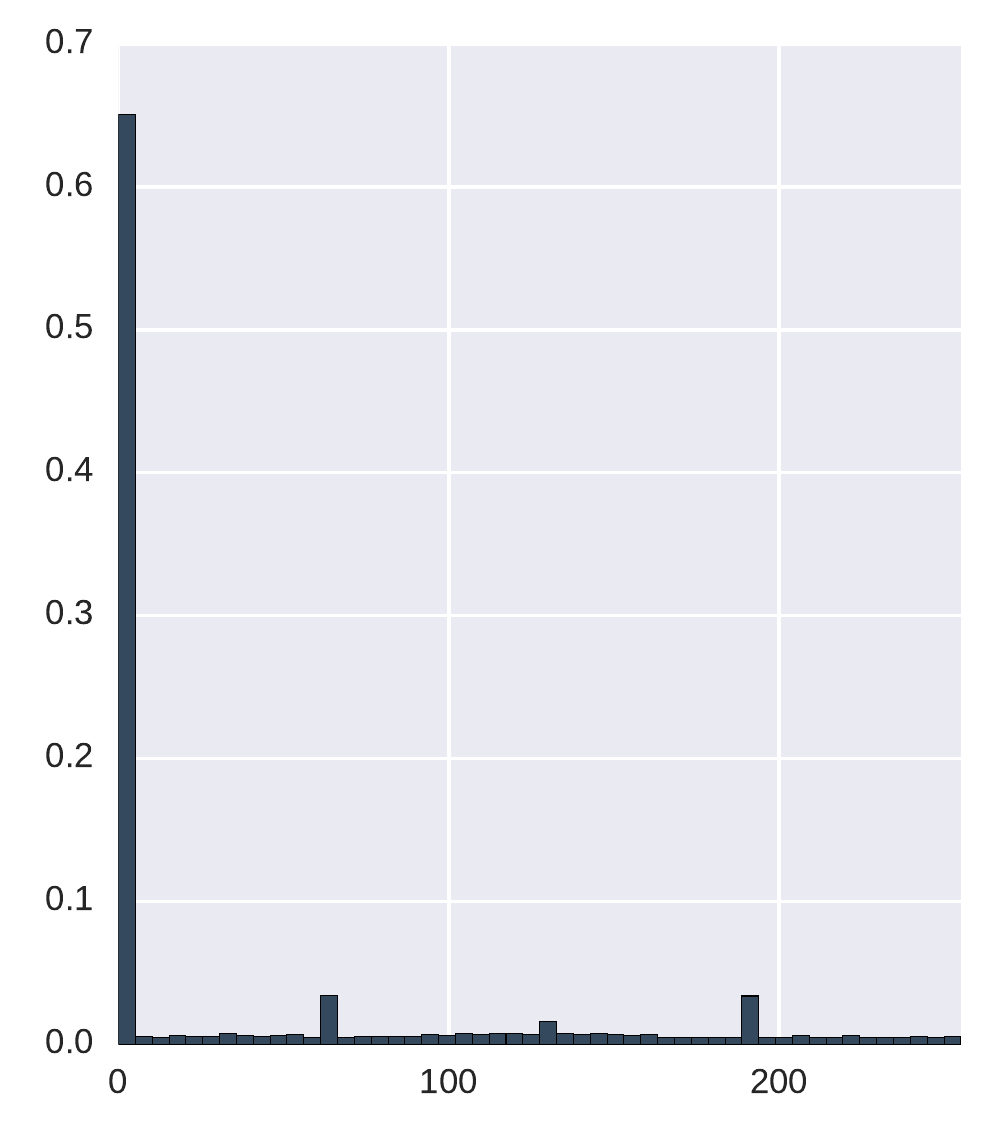}} \hfill
	\subfloat[]{\includegraphics[width=0.22\textwidth]{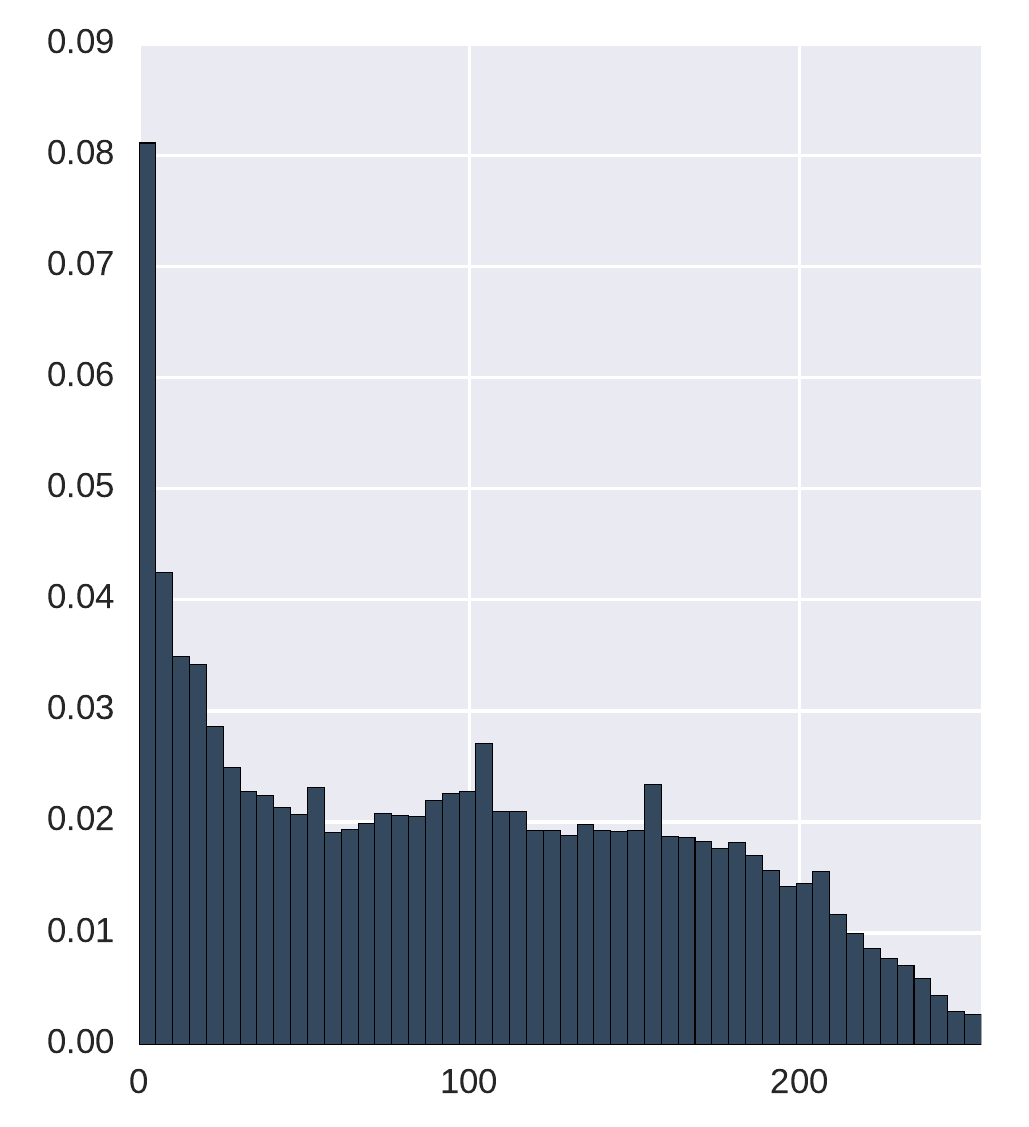}} \hfill
	\subfloat[]{\includegraphics[width=0.22\textwidth]{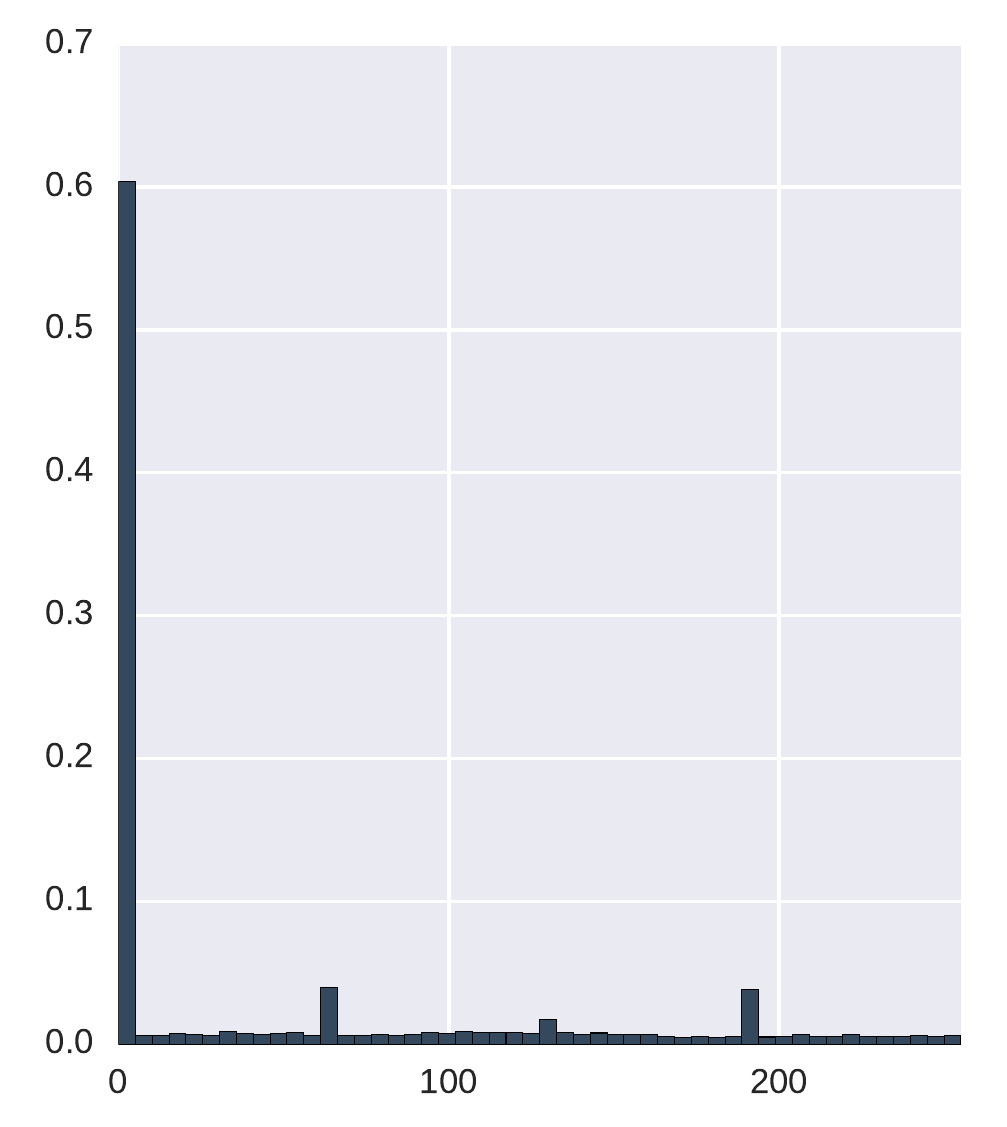}} 
	\caption{\textbf{Ciphertext sensitivity}: (a),(b) plaintext distribution of the Lena image and corresponding ciphertext distribution; (c),(d) plaintext distribution of the cat image and corresponding ciphertext distribution. Since the ciphertext distributions are almost identical, frequency analysis is difficult and the system is robust against ciphertext-only attacks.}
	\label{distribution}
\end{figure}

Known-plaintext attacks are ones where an attacker has access to an example of a plaintext (a message) and a corresponding ciphertext (a weight matrix $\boldsymbol{W}^{o}$) and attempts to crack the secret key via a comparative analysis of changes between them. For instance, by analyzing changes in the ciphertexts of images which differ by just a few pixels, it might possible to obtain part of the mapping involved in encryption. Figures~\ref{plaintext_sensitivity}(a) and (f) show original images and Figs.~\ref{plaintext_sensitivity}(c) and (h) show slightly distorted versions where 1\% of the pixels were randomly changed. The corresponding encrypted images (matrices $\boldsymbol{W}^{o}$) are visualized in Fig.~\ref{plaintext_sensitivity}(b), (g), (d), and (i). Only small changes in the plaintext led to considerable changes in the chiphertext; these differences are visualized in Fig.~~\ref{plaintext_sensitivity}(e) and (j) and amount to about $35\%$. Thus, our system is sensitive to slight modification of the plaintext and therefore renders known-plaintext attacks very difficult. 

In ciphertext-only attacks, an attacker has access to a set of ciphertexts, however has some knowledge about statistical distribution of plaintexts. Using frequency analysis of ciphertexts, for instance, exploiting the fact that ``e'' is the most frequent character in English texts, one can map the most frequent parts in a ciphertext to corresponding plaintexts. Figure~\ref{distribution} shows frequency distributions for the plaintexts and ciphertexts of the images ``Lena'' and ``cat''. Although the plaintext distributions of two images differ, their ciphertext distributions are very similar. From these distributions it is evident that most of the elements ($\approx 50\%$) in the ciphertext ($\boldsymbol{W}^{o}$) are zero and that the non-zero elements are uniformly distributed. Thus, frequency analysis will be ineffective and the proposed system is robust against ciphertext-only attacks.

According to Shannon \cite{shannon1949communication}, diffusion and confusion are the two fundamental properties of a good cryptography system. A system that has the diffusion property is one where a small change in either plaintext or key causes a large change in the ciphertext. A system with the confusion property is one where the mapping between plaintext and ciphertext is complex. Our experimental results indicate that the proposed system has both these properties.

\begin{figure}[t!]
\centering
\begin{minipage}{0.45\textwidth}
	\includegraphics[width=1.0\textwidth]{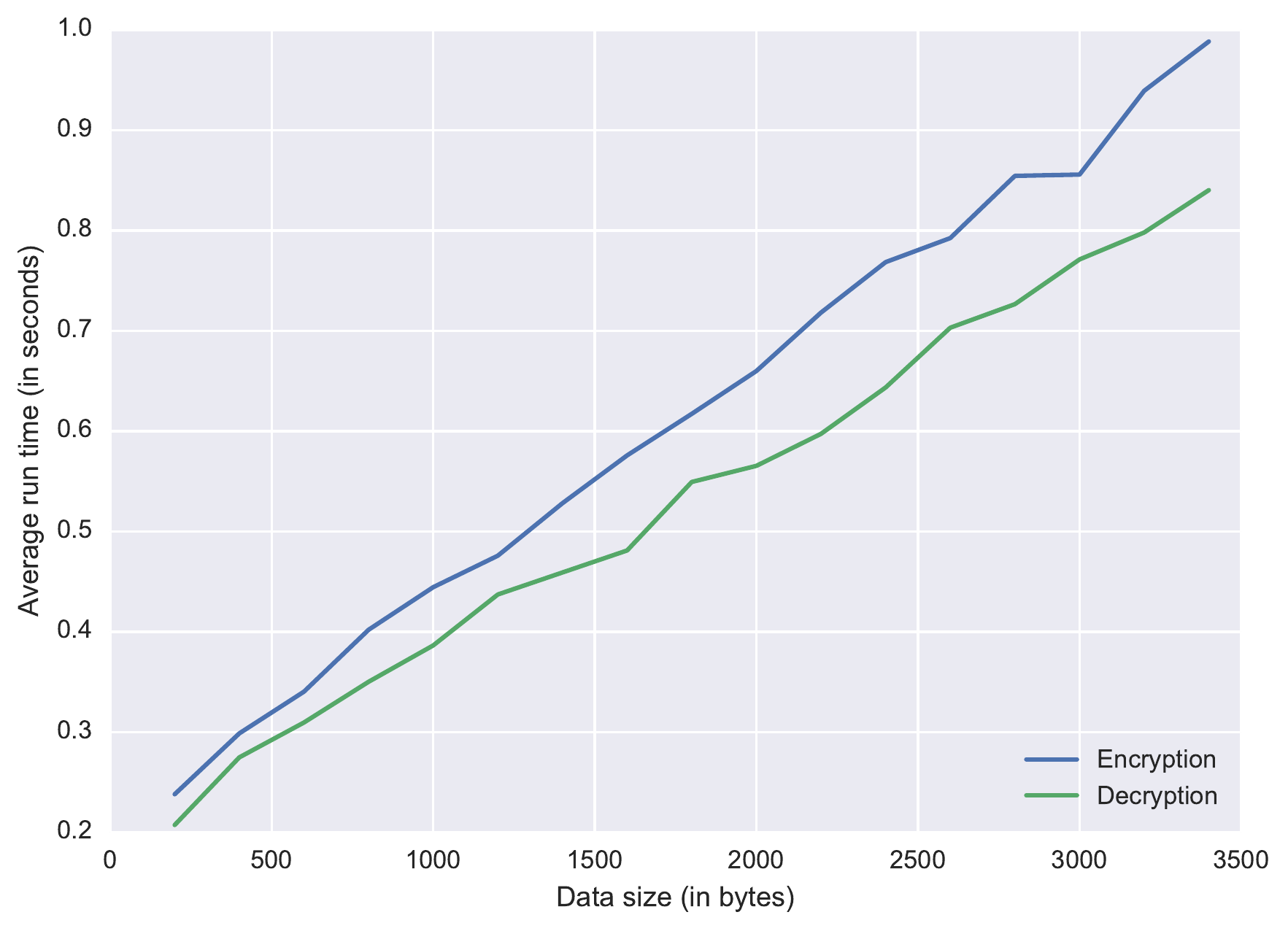}
\end{minipage}
\begin{minipage}{0.4\textwidth}
		\begin{tabular}{ll}  
			\toprule
			environment \\
			\midrule
			processor & 2.7 GHz Intel Core i5\\ 
			memory &  8 GB\\
			OS & OS X El Capitan\\
			language & Python 3.5, Numpy\\
			\bottomrule
		\end{tabular}
\end{minipage}
	\caption{Run times for encryption and decryption for different message sizes.}
	\label{runtime}
\end{figure}

\subsection{Performance}

To evaluate the runtime performance of our proposed system, we determined average encryption and decryption times for messages of different sizes. Our results are shown in Fig.~\ref{runtime}. For instance, encrypting and decrypting a 3KB message took less than one second each and runtimes were found to increase linearly with the message size. Our approach therefore scales well and can be used in real-time applications.

\section{Conclusion}\label{sec:conclusion}

In this paper, we proposed a novel neural cryptography scheme based on the capability of echo state networks to memorize and reproduce sequences of input data. The proposed system was found to be robust against common security attacks and satisfies the fundamental cryptographic properties of diffusion and confusion. Moreover, our approach is scalable, suitable for real-time applications, and does not require special purpose hardware (such as GPUs) for its computations.

\bibliographystyle{splncs03}
\bibliography{refs}

\end{document}